# From customer to product: design tools for the visually impaired

Eduardo Augusto Monteiro de Almeida[a,b,*], Guillaume Thomann[b], Angelina Dias Leão Costa[a]

[a]*Univ. Federal da Paraíba, Campus I - Cidade Universitária - Centro de Tecnologia, LACESSE, Bloco N, João Pessoa - 58051-900, Brazil*
[b]*Univ. Grenoble Alpes, CNRS, Grenoble INP, Institute of Engineering Univ. Grenoble Alpes, G-SCOP, 38000 Grenoble, France*

\* Corresponding author. Tel.: +55-83-99931-3050; Tel.: +33-07-64-53-18-18. *E-mail address:* eduardo-augusto.monteiro-de-almeida@grenoble-inp.fr

**Abstract**

Navigation in new or unknown environments is vital, especially for visually impaired individuals. While many solutions exist, few are tailored to specific disabilities, often due to limited collaboration with handicap users in the design process. This article examines 7 tools that enable visually impaired users to participate in design, selected through a systematic review and analyzed for affinities, differences, and applications. The study suggests correlations among the tools, offering a foundation for a methodology that enhances inclusive design and accessibility.

## 1. Introduction

The experience of visually impaired people in the architectural design process is little explored. These people use their remaining senses to navigate through space, a process called wayfinding. Including these people in the development of projects results in more sensitive and effective solutions, although there is currently a distance between them and designers. [7, 34, 40, 53]

Traditional methodologies fail to include VIP in the design process. The HCD approach and participatory projects mitigate this issue by directly involving end users. Interaction and dialogue with these people are essential to understanding their needs, promoting accessibility in PD [3, 35].

This paper aims to investigate participatory tools used to integrate people with visual impairments into design. The guiding question is: How does the association of these tools enhance the capabilities of people with visual impairments in the design process?

The methodology consists of three phases: A systematic review identifies 38 studies that present approaches allowing the collaboration of blind people in the design process; A comparative analysis examines the tools found and the potential of their application; A matrix of discoveries structures the findings, presenting the frequency of application and the possibilities of implementation.

The systematization of the collected information demonstrates that it is possible to include blind people in the design process, which contributes to producing artifacts of perception and spatial orientation in built spaces..

The discussion of this participatory approach with the blind person as the protagonist favors a more universal didactic practice for teaching design, highlighting the potential of the user, and not their limitations, as is customary.

**Nomenclature**

HCD     Human-Centered Design
VIP      Visually Impaired Person
PD      Participatory Design

## 2. Theoretical Framework

### 2.1. Human-Centered Design

Human-Centered Design (HCD) begins with the analysis of users' needs, desires, and behaviors, seeking solutions that are desirable, feasible, and viable. The HCD methodology consists of three stages: Listen, Create, and Implement, and promotes greater dynamism in product development. However, there is a growing effort to integrate human particularities, focusing on emotional factors and placing the user at the center of each phase of the project. This is crucial to creating more accessible environments and meeting users' real needs, reducing the distance between designers and end recipients. [31, 36]

### 2.2. User Experience Design

The User Experience Design (UXD) approach is an iterative cycle with Analysis, Design, Prototyping, and Evaluation phases, allowing the active participation of users as co-creators under the supervision of the designer. The UXD process encompasses several areas, including Interaction Design, Software Engineering, Product Engineering, Information Architecture, and Visual Design. User experience is seen as an interaction between the individual and the environment, and it is essential to consider it as a priority in the design of interactive products. Without a clear understanding of the experience, interactive products cannot shape appropriate experiences. [18, 22, 23]

### 2.3. Visually Impaired User

Visual impairment results from ocular or cortical disorders and is classified into four levels: normal vision, moderate visual impairment, severe visual impairment, and blindness. Blind people do not have sufficient vision to read ink and need to use other senses such as touch, hearing, smell, taste, and kinesthetics. People with low vision use their residual visual potential to explore the environment and learn. The limitation is not in the impairment but in the environment and in the societal understanding of visual impairment, which remains insufficient, reflecting in low expectations about the intellectual performance of these people.

Understanding the environment and its spatial organization is crucial for the efficient mobility of people with visual impairment. They construct concepts using alternative pathways such as touch, hearing, and smell. Touch, in particular, is essential for spatial understanding and should be developed as a critical skill, not neglected. Tools that transmit tactile information are vital. Inclusive solutions are necessary so that all people can live together in society in a safe and autonomous manner, considering the user when making design decisions and adopting information systems that are understandable to all. [5, 11, 20]

### 2.4. Wayfinding System

Spatial orientation is a cognitive process that involves two skills: spatial orientation and wayfinding. Spatial orientation is the ability to understand where one is, while wayfinding is the ability to move around by performing mental activities such as information processing, decision-making, and execution. These skills are complementary and depend on the information contained in the environment and the individual's ability to perceive and process this information, using mental maps during movement. [9, 39]

Wayfinding is a systematic movement process based on information captured by sensory systems such as visual, auditory, tactile, and olfactory. Effective communication between the building and its users depends on various solutions, such as tactile, auditory, pictorial, chromatic, and alphanumeric resources. Architectural experiences resulting from sensory perception must be guaranteed by designers for efficient spatial orientation. Even with technological advances, sensory references remain essential in wayfinding. [6, 7, 8]

In built spaces, architectural and urban elements serve as references for spatial orientation. The shape, color, and function of the building help create mental images, which are essential for locomotion. Spatial accessibility goes beyond simply entering or reaching a location; it involves ensuring that anyone can move around easily. It is crucial that complex environments such as hospitals and airports ensure legibility and consider individual factors of users, such as emotional state and previous experiences. [4, 16, 28, 48]

Environmental information is transmitted by architectural elements or characteristics of spaces used as references for orientation. These elements are identified by color, texture, shape, smell, and noise. These characteristics are essential for individuals to obtain information and move to their destinations. [28, 38]

## 3. Method

The methodology is structured in three phases: results of a systematic review that identifies 38 studies presenting approaches that enable the collaboration of blind people in the design process; comparative analysis between the tools identified and their potential applications; and the structuring of a findings matrix, which synthetically presents the frequency and possibilities of application.

### 3.1. Systematic Review

Systematic reviews are particularly useful for integrating information from a set of studies conducted separately on a given area of knowledge. These studies may present conflicting or coinciding results, while also identifying topics that require further evidence, thereby guiding future research [43]. As the first methodological phase, the literature review feeds all subsequent stages by presenting the main concepts and debates on the topic [1].

### 3.2. Comparative analysis

The comparative analysis stage evaluates the tools identified in the systematic review based on criteria such as ease of use and impact on the autonomy of blind people. The comparative method allows for the observation and contrast of the characteristics of each approach, following inclusive design guidelines and best practices in accessibility. As highlighted by [30], this method is essential for "identifying patterns and understanding variations" between different solutions.

*3.3. Findings Matrix*

The findings matrix serves as an analytical tool for identifying and visually communicating findings, such as adaptations resulting from design flaws, user misunderstandings, and daily operational difficulties. Adapted for this research, it facilitates the reading and understanding of results, transforming it from a mere record of problems into an analytical instrument. Classifying information for comparison simplifies the identification of the origins of problems. Originating from Post-Occupancy Assessment studies, the graphic matrix summarizes the main findings, promoting an integrated and non-fragmented analysis of the studied theory [42].

## 4. Results and Discussions

*4.1. Results of the Systematic Review*

The main search sources for this research were CAPES Journal, Google Scholar, ScienceDirect, Theses.fr, and the Brazilian Digital Library of Theses and Dissertations. The most relevant results were those addressing methodology, theoretical basis, and systematic review. Initially, the searches yielded 4,155 documents, which underwent a rigorous selection and filtering process, culminating in 38 articles.

The filtering process included the selection of keywords, which were grouped and inserted into the databases. Inclusion and exclusion criteria were adopted, such as the period between 2018 and 2023, peer-reviewed and open-access journals, and availability online in any language. Duplicate files were eliminated, reducing the number of documents to 3,383.

Next, documents were eliminated based on titles, discarding 3,026 that were not related to the research objectives, leaving 357. After verifying open access to each work, 89 documents were excluded due to restricted access, resulting in 268 documents. Reading the abstracts finally selected the 38 most relevant works for the research.

The selected works mainly address inclusion and accessibility through the development of technologies. The diversity of approaches and methodologies is evident among the analyzed studies. Some highlight the importance of co-design, where visually impaired users actively participate in the design process, demonstrating that user participation can significantly improve the usability and effectiveness of the developed products.

The application of 3D technologies is recurrent, especially in research focused on exhibition spaces such as museums. These studies demonstrate how the creation of adapted resources and the digital fabrication of objects can make cultural environments more accessible. Additionally, some research focuses on creating educational and multisensory experiences for children with visual impairments, while others evaluate usability and user experience (UX) to develop intuitive and effective interfaces for these users.

Although co-design is a common approach among studies, methodologies vary from more technical and formal approaches to more creative and participatory methodologies. The contexts in which technologies and methods are applied also differ. Some studies focus on educational environments and museums, while others explore consumer products.

Certain trends stand out, such as the integration of emerging technologies, including 3D printing and digital tools. The expansion of the co-design concept, involving not only visually impaired users but also other stakeholders—such as designers, engineers, and educators—is highlighted as a factor contributing to the development of technologies that are not only functional but also enjoyable and easy to use.

*4.2. Notes from Comparative Analysis*

The comparative analysis of the selected studies reveals a set of recurrent tools among the authors, highlighting trends and areas of interest in accessibility and assistive technology research. Among the most used methodological tools are interviews, focus groups, workshops, usability tests, 3D modeling/prototyping, questionnaires, and guided tours, each supported by a base of authors who have contributed to advancing investigative practices in these areas.

Interviews stand out as a tool used in studies requiring direct interaction between researchers and participants, as seen in the studies by [2, 12, 57]. These authors investigated aspects related to the usability of technologies and the development of systems adapted to the needs of people with disabilities. Interviews provide a deep understanding of the user experience and are valuable for capturing qualitative and subjective perceptions.

Focus groups, employed by authors such as [3, 32], serve as a tool that fosters interaction among participants to discuss specific topics. This methodology is effective in generating collective insights, particularly in studies on assistive technologies and inclusive design.

Workshops, used by authors such as [24, 27], create a collaborative environment where solutions are co-created with users. This method is especially useful in participatory design projects, allowing participants to actively contribute to the development of products and systems.

Usability testing, applied by authors such as [15, 19], assesses the efficiency and functionality of assistive technologies. This tool aims to identify potential improvements in systems to ensure an accessible and intuitive user experience.

3D modeling and prototyping, utilized by authors such as [13, 33], stand out for enabling the creation of tactile representations of objects and spaces. This approach is particularly useful in studies focused on accessibility for visually impaired individuals, allowing for the simulation and adaptation of environments before final implementation, ensuring a prior assessment of accessibility. Questionnaires, in turn, appear in studies such as those by [17, 47] as a valuable quantitative tool for large-scale data collection, evaluating users' and experts' perceptions of accessibility technologies and practices.

The guided tour, a more immersive methodology, is used by authors such as [3, 35] to conduct a contextualized analysis of users' interactions with the environment, identifying accessibility barriers and facilitators in daily life.

*4.3. Structuring the Discovery Matrix*

Table 1 presents the findings matrix, identifying seven main tools from the systematic review: interview, focus group, workshop, usability testing, 3D modeling/prototyping, questionnaire, and guided tour.

Each tool is accompanied by its frequency of application in the reviewed studies and the respective references of the authors who employed them. The first column lists the identified tools, the second indicates the number of times each tool was applied in the reviewed studies, and the third cites the corresponding authors.

The frequencies reflect the application of methods by different authors, with some utilizing multiple tools in their studies. This explains why the total sum of frequencies exceeds the number of authors and why certain authors appear associated with multiple tools.

Table 1. Discovery Matrix

| Tools | Frequency | Authors |
| --- | --- | --- |
| Interview | 19 | [1]; [2]; [12]; [13]; [17]; [21]; [25]; [26]; [32]; [35]; [33]; [37]; [46]; [49]; [50]; [51]; [52]; [55]; [57]. |
| Focus Group | 6 | [2]; [32]; [35]; [33]; [34]; [37]. |
| Workshop | 6 | [2]; [24]; [26]; [27]; [32]; [37]. |
| Usability Testing | 5 | [2]; [10]; [15]; [19]; [54]. |
| 3D Modeling/ Prototyping | 10 | [13]; [14]; [25]; [26]; [29]; [30]; [33]; [34]; [45]; [54]. |
| Questionnaire | 9 | [17]; [29]; [41]; [47]; [50]; [51]; [52]; [55]; [56]. |
| Guided Tour | 3 | [2]; [17]; [35]. |

The structuring of the results matrix based on the systematic review enabled the visualization of the frequency of each approach and the combination of different methodological strategies to enhance the inclusion of people with visual impairments in the design process. The identified tools, when integrated, can provide more robust results and ensure that the developed solutions are effective, functional, and user-centered.

One possible combination involves interviews, workshops, and usability tests. Interviews allow for an in-depth exploration of users' individual needs and expectations, offering essential qualitative insights into their specific requirements. However, interviews alone may not be sufficient to generate practical solutions. In this context, workshops foster a collaborative environment in which users and designers work together to develop more suitable and personalized solutions. At the end of this process, usability tests assess the effectiveness of the proposed solutions.

Another relevant combination includes 3D modeling, workshops, and guided tours. 3D modeling is particularly useful in projects involving physical accessibility, such as those focused on public or cultural spaces. The creation of tactile prototypes enables people with visual impairments to better understand a space before interacting with it directly. Based on this representation, workshops allow users to actively participate in the creation and refinement of models, adjusting them to their needs. Subsequently, guided tours test these solutions in real contexts, offering a practical assessment of how the adaptations function in everyday life.

The combination of 3D modeling, usability testing, and interviews is also significant in projects that involve the development of complex objects or environments. 3D modeling facilitates visualization and tactile interaction with objects, while usability testing verifies whether the proposed solutions effectively meet the needs of visually impaired users. Interviews complement this process by providing subjective feedback on the user experience, which is crucial for refining the final design.

Lastly, integrating workshops, interviews, and guided tours enables a practical evaluation of solutions in real-world scenarios. Workshops encourage co-creation with users, interviews deepen the understanding of their individual needs, and guided tours assess the functionality of solutions in daily contexts, ensuring they align with users' real-life demands.

Thus, the integration of these methodologies enhances the inclusive design process, ensuring that solutions are not only technically accessible but also intuitive, engaging, and tailored to the actual needs of visually impaired users.

## 5. Considerations

This article demonstrated, based on the reviewed studies, a series of advances and insights in the field of Participatory Design. The structuring of the results matrix through the systematic review enabled the visualization of the frequency of each approach and the combination of different methodological strategies to enhance the inclusion of people with visual impairments in the design process. The identified tools, when integrated, can provide more robust results and ensure that the developed solutions are effective, functional, and user-centered.

The systematic review conducted in this study highlights a strong commitment to inclusion and accessibility in the design of technologies and products for people with disabilities. The similarities among the analyzed research, particularly the focus on co-design and the application of technologies such as 3D printing, point to a collective trend toward more participatory design practices. However, methodological differences emphasize the need for personalized approaches that consider the specificities of each application and context.

By organizing key findings from the analyzed studies into a Findings Matrix, the correlations between works became evident. While some studies adopt co-design to actively involve people with visual impairments in the design process, others utilize technologies such as 3D printing, digital simulations, and tactile models to develop more inclusive and adaptable solutions. Despite methodological variations, all studies underscore the importance of improving usability and user experience, ensuring that products and environments are designed based on user involvement and interaction.

The integration of these methodologies enriches the inclusive design process, ensuring that solutions are not only technically accessible but also intuitive, engaging, and adapted to the real needs of visually impaired users. This diversity of approaches demonstrates the complexity and broad scope of the field of design.